\documentclass[10pt]{iopart}

\usepackage{graphicx}
\usepackage{dcolumn}
\usepackage{bm}
\usepackage{subfigure}
\usepackage{hyperref}
\usepackage{color}
\usepackage{amssymb}
\newcommand{\jupyter}{\textsc{jupyter} }
\newcommand{\alexnet}{\textsc{alexnet} }
\newcommand{\resnext}{\textsc{resnext} }
\newcommand{\imagenet}{\textsc{imagenet} }
\newcommand{\rem}[1]{}

\begin{document}


\title{Interpretable, calibrated neural networks for analysis and understanding of inelastic neutron scattering data}

\rem{
\author{Keith T. Butler}
\affiliation{SciML, Scientific Computing Department, STFC Rutherford Appleton Laboratory, Harwell Campus, Didcot, OX11 0QX, UK}
\affiliation{%
Department of Materials Science and Engineering,
University of Oxford,
21 Banbury Rd, 
Oxford OX2 6HT
}
\email{keith.butler@stfc.ac.uk}

\author{Manh Duc Le}%
\affiliation{ISIS Neutron and Muon Source, STFC Rutherford Appleton Laboratory, Harwell Campus, Didcot, OX11 0QX, UK}

\author{Jeyan Thiyagalingam}
\affiliation{%
 SciML, Scientific Computing Department, STFC Rutherford Appleton Laboratory, Harwell Campus, Didcot, OX11 0QX, UK\\
}
\affiliation{%
Department of Engineering Science,
University of Oxford,
Parks Road,
Oxford, OX1 3PJ
}

\author{Toby G. Perring}%
\affiliation{ISIS Neutron and Muon Source, STFC Rutherford Appleton Laboratory, Harwell Campus, Didcot, OX11 0QX, UK}
}

\author{Keith T. Butler$^{1,2}$\footnote{email: keith.butler@stfc.ac.uk}, Manh Duc Le$^3$, Jeyan Thiyagalingam$^{1,4}$, Toby G. Perring$^3$}
\address{$^1$SciML, Scientific Computing Department, STFC Rutherford Appleton Laboratory, Harwell Campus, Didcot, OX11 0QX, UK}
\address{$^2$Department of Materials Science and Engineering, University of Oxford, 21 Banbury Rd, Oxford OX2 6HT}
\address{$^3$ISIS Neutron and Muon Source, STFC Rutherford Appleton Laboratory, Harwell Campus, Didcot, OX11 0QX, UK}
\address{$^4$Department of Engineering Science, University of Oxford, Parks Road, Oxford, OX1 3PJ}

\date{\today}

\begin{abstract}
Deep neural networks provide flexible frameworks for learning data representations and functions relating data to other properties and are often claimed to achieve `super-human' performance in inferring relationships between input data and desired property. In the context of inelastic neutron scattering experiments, however, as in many other scientific scenarios, a number of issues arise: (i) scarcity of labelled experimental data, (ii) lack of uncertainty quantification on results, and (iii) lack of interpretability of the deep neural networks. In this work we examine approaches to all three issues. We use simulated data to train a deep neural network to distinguish between two possible magnetic exchange models of a half-doped manganite. We apply the recently developed deterministic uncertainty quantification method to provide error estimates for the classification, demonstrating in the process how important realistic representations of instrument resolution in the training data are for reliable estimates on experimental data. Finally we use class activation maps to determine which regions of the spectra are most important for the final classification result reached by the network.

\end{abstract}

\maketitle
\ioptwocol


\section{\label{sec:intro} Introduction}

Neural networks have witnessed a renaissance in the last decade, dramatically improving on existing state of the art performance in fields from image processing to automatic language translation. In the sciences, it has been proposed that we are on the cusp of a ``fourth paradigm'' of data-driven discovery~\cite{hey2009fourth, agrawal2016perspective}, an idea that is supported by the recent explosion in research using machine learning (ML), not least in materials science and condensed matter physics~\cite{carrasquilla2017machine, butler2018machine, bednik2019probing, morita2020modelling}.

Experiments at large-scale facilities such as neutron sources would seem to be a natural fit for the application of data-driven methods for analysis~\cite{hey2020machine}. Indeed there have been a number of recent publications applying machine learning to many aspects of neutron science, from pulse discrimination in scintillator detectors~\cite{doucet2020machine}, to enhancing resolution~\cite{islam2019} and exploration of collected data~\cite{hui2019volumetric}. ML techniques have been used for the analysis of diffuse neutron scattering in spin-ice systems~\cite{samarakoon2020}, small angle neutron scattering~\cite{archibald2020classifying, demerdash2019using} and for constructing inter-atomic potentials to compare to inelastic neutron scattering data~\cite{qian2018temperature}. However, questions still remain to be answered before machine learning becomes a fully integrated and trusted part of the neutron data analysis workflow; for example can we place uncertainty estimates on our predictions and can we explain why a neural network predicted what it did? A particularly challenging case is the interpretation of inelastic neutron scattering (INS) data from single crystal samples. 

INS is a powerful technique to investigate the wave vector and frequency (equivalently excitation energy) dependency of the spectrum of excitations in condensed matter.
The neutron scattering cross-section is particularly simple: it directly yields two-particle correlation functions, for example the particle-particle correlation function, or in the case of magnetic systems the spin-spin correlation function, and by virtue of the fluctuation-dissipation theorem, the generalised response functions of the system~\cite{lovesey1984, squires1978}. Consequently INS has played a central role determining, for example, phonon and spin wave dispersion relations and the spectra of magnetic fluctuations in materials, which in turn can play central roles in the thermal and charge transport properties, and the mechanisms of exotic phenomena such as high temperature superconductivity and heavy fermion behaviour~\cite{chen2019anisotropic, wang2013doping, kieslich2018hydrogen, li2020ultralow}.

The richest information comes from experiments with single crystal samples, where the spectra can be explored as a function of all four components of wave vector and frequency. The latest INS spectrometers for such experiments at pulsed neutron sources (e.g.~\cite{LET, ARCS, 4SEASONS}) and equivalent instruments at reactors (e.g.~\cite{IN5}) allow the incident neutron energy to be fixed, and typically have $\approx$3~steradians of position sensitive area detectors divided into $\sim 10^5$ detector elements, in each of which the energy of every scattered neutron (and hence the energy transferred to excitations) is resolved into $\sim 200 - 500$ channels. With just one crystal orientation, therefore, data are collected in $>10^7$ voxels in a three dimensional (3D) space (corresponding to the the two coordinates on the area detector, and energy transfer). In the case of quasi-2D magnetic materials, where the magnetic coupling is significant only within planes of atoms but negligible between planes, these three coordinates can be transformed into those of wave vector within the plane and energy transfer, thereby in principle enabling the entire excitation spectrum to be measured in parallel. An extension of this approach is to combine $\sim 100 - 500$ measurements between which the crystal is rotated by a few tenths of a degree. The crystal rotation provides another independent coordinate so that the full 4D wave vector - excitation energy space of excitations can be measured, in $\sim 10^9 - \sim 10^{10}$ voxels (\cite{horacepaper} and references therein).

The sizes of the datasets collected on these instruments are challenging to analyse and interpret with conventional but well-established and well-understood methods and software. The motivation for this work has been to investigate to what extent ML methods can reduce the labour of inelastic neutron data analysis while still returning reliable and interpretable information. In this contribution we re-examine the data from a well-understood INS measurement of spin waves in a single crystal of Pr(Ca$_{0.9}$Sr$_{0.1}$)$_2$Mn$_2$O$_7$~\cite{johnstone2012}, a moderately complex magnetic material, using recent advances in deep learning. Different models of the exchange interactions in Pr(Ca,Sr)$_2$Mn$_2$O$_7$ were proposed based on the observed magnetic and crystallographic structure, but INS measurements of the spin wave spectrum and theoretical modelling were key to distinguishing between them. We show that ML techniques can be used to distinguish between the models and can highlight the diagnostic regions of the spectrum responsible for this categorisation.

In Section~\ref{sec:pcsmo} we briefly describe the interest in and theoretical background to the physics of the manganites, and Pr(Ca$_{0.9}$Sr$_{0.1}$)$_2$Mn$_2$O$_7$ in particular, together with the different exchange models. In Section~\ref{sec:data} we discuss the experimental data and how it was modelled using linear spin wave theory to create training datasets for our neural networks, taking account of the instrument resolution. In Section~\ref{sec:nn}, the main part of the paper, we introduce the neural network classifier, incorporate uncertainty quantification into the network and show how class activation maps may be used to understand its inferences. Finally, we discuss future challenges and summarise our conclusions.

\section{Exchange interaction models in half-doped manganites} \label{sec:pcsmo}

Perovskite manganese oxides RE$_{(1-x)}$A$_x$MnO$_3$ (RE = rare earth, A = Ca,Sr,Ba,Pb) and their layered analogues have been widely studied not just because of the colossal magnetoresistance (CMR) they can show - up to many orders of magnitude change in resistivity in applied magnetic fields of a few Tesla - but also because of the rich physics that arises from the coupling of the charge, spin, lattice and orbital degrees of freedom~\cite{tokura2006manganites}. Typically the most pronounced CMR is with light hole doping $x\approx 0.2-0.4$ where the material is balanced between being in a ferromagnetic (FM) metallic phase and an antiferromagnetic (AFM) charge and orbital ordered (CO) insulating phase. One of the commonest AFM CO phases is found near half doping, $x=0.5$, and the nature of this phase was the topic of the INS study in ref.~\cite{johnstone2012}.

The structure of the perovskite manganese oxides consists of manganese ions, each surrounded by six oxygen ions forming a distorted octahedron, and these octahedra are in turn connected at their vertices to create a (nearly) cubic network, with a rare earth or alkaline earth cation filling the gap between the octahedra. Pr(Ca,Sr)$_2$Mn$_2$O$_7$ is a half-doped bilayer manganite consisting of pairs of MnO$_6$ octahedral layers, which are separated by a layer of the interstitial atoms that reduce the magnetic exchange coupling between bilayers by two orders of magnitude compared to the intra-bilayer couplings. 

\begin{figure}
\centering
    \includegraphics[width=\columnwidth]{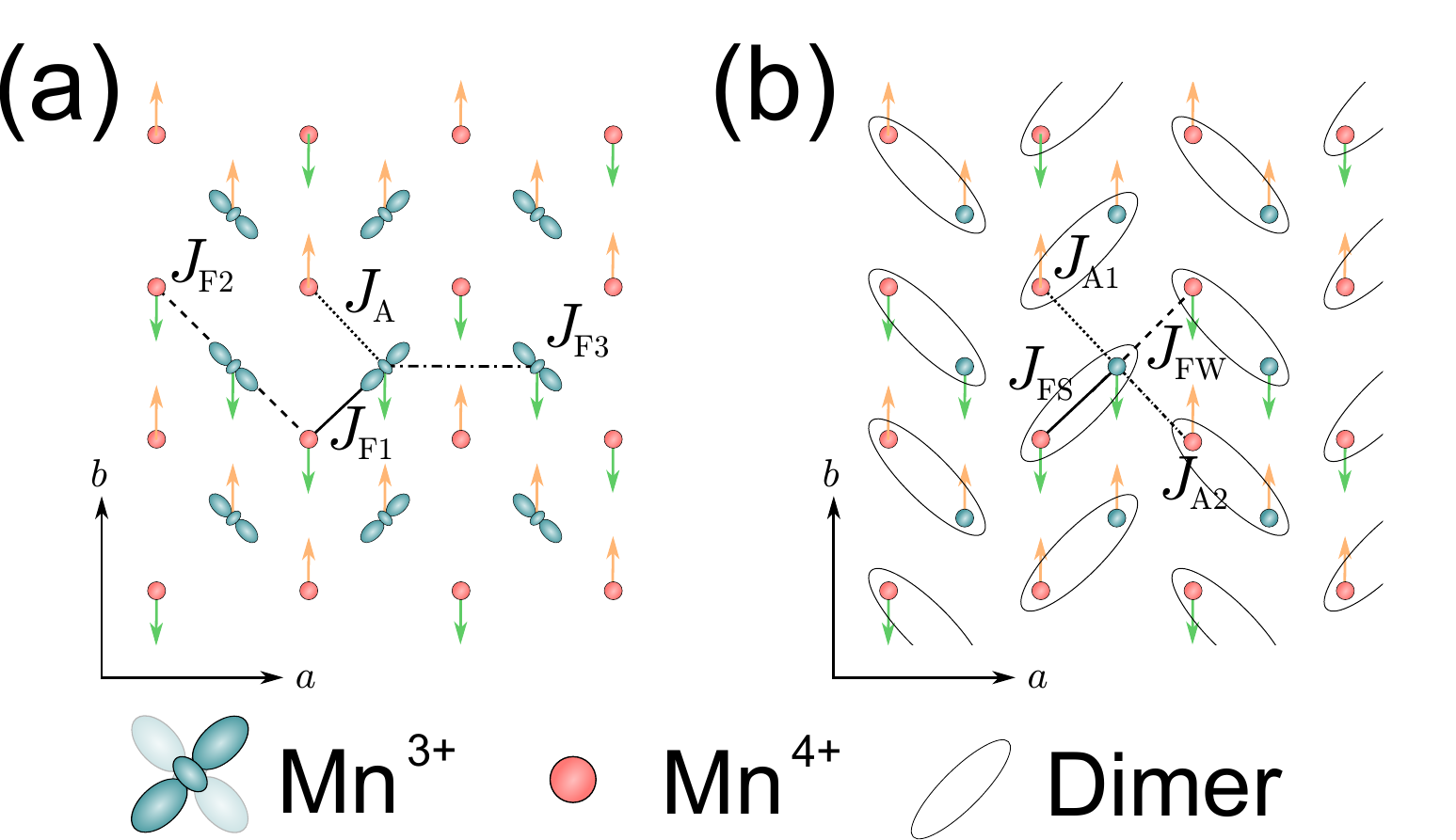}
\caption{\label{fig:model} Two magnetic exchange models in a single sheet of Mn ions in a half-doped manganite showing the CE-type magnetic ordering. (a) Goodenough model (b) dimer model, similar to Goodenough but neighbouring Mn$^{3+}$/Mn$^{4+}$ sites along the ferromagnetic zig-zag chains are loosely bound in a dimer.}
\end{figure}

The famous and long-standing \emph{Goodenough} model for the AFM CO phase is shown in Figure~\ref{fig:model}(a). It is based on the Goodenough-Kanamori-Anderson rules~\cite{goodenough1955, kanamori1959} for magnetic superexchange together with a checkerboard ordering of equal numbers of formally $3+$ and $4+$ valence Mn ions. The octahedra around the Mn$^{3+}$ are elongated due to the singly occupied higher energy $e_g$ orbitals. Accommodating this elongation within the structure results in a herringbone pattern of the $e_g$ orbitals and the experimentally observed CE-type magnetic structure, consisting of ferromagnetic ordering along zig-zag chains with antiferromagnetic ordering between neighbouring zig-zags.

A very different and somewhat controversial model came to prominence in 2002, following a seminal neutron diffraction experiment~\cite{daoudaladine2002}. In this model, known as the Zener polaron model, adjacent Mn$^{3+}$ and Mn$^{4+}$ spins form dimers with strong ferromagnetic intra-dimer coupling, but much weaker inter-dimer coupling (Figure~\ref{fig:model}(b)). The two models have the same underlying periodicities and it is difficult to distinguish between them using diffraction techniques as this relies on the presence or absence of (and the differences between the intensities of) weak superlattice Bragg peaks. However, measuring the spin wave spectrum offers a way to distinguish between the models. The limiting case of strong intra-dimer coupling was straightforwardly eliminated on the grounds of the periodicities of the spin wave dispersion relations~\cite{johnstone2012} compared to the data. A more realistic scenario is that of weaker dimerisation.
This \emph{dimer} model was eventually eliminated in ref.~\cite{johnstone2012}, but it was far from straightforward, because with an appropriate set of magnetic exchange parameters the dimer model can closely reproduce the spin wave dispersion relations in the Goodenough model, with only subtle differences in the intensities throughout most of the Brillouin zone. Determining which of the Goodenough and dimer models correctly described the magnetic interactions was the primary result of ref.~\cite{johnstone2012} which concluded that the Goodenough model best fits the INS data.

Both the Goodenough and dimer models can be described by a spin Hamiltonian taking the standard Heisenberg form

\begin{equation}
    \mathcal{H} = \sum_{i,j} J_{ij} \mathbf{S}_i\cdot\mathbf{S}_j + D \sum_i (S_i^z)^2
\end{equation}

\noindent where the atom pairs $ij$ are not the same in the two models, so the topology of the exchange interactions $J_{ij}$ will differ. In this formulation, a positive $J_{ij}$ indicates antiferromagnetic interactions (this is opposite to the convention used in ref~\cite{johnstone2012}). The final term is a single-ion anisotropy term which with $D>0$ tends to keep the spins in the $a-b$ plane consistent with observations. Figure~\ref{fig:model} shows the exchange topology of each model, and Table~\ref{tab:tab1} summarises the published exchange interactions~\cite{johnstone2012} deduced for the two models.
 
In the Goodenough model the nearest-neighbour FM interaction $J_{F1}$ along the zig-zags is supplemented by two next-nearest-neighbour interactions $J_{F2}$ and $J_{F3}$, which need not be FM but must be permitted to reproduce experimentally observed periodicities in the experimental data. In addition there is an AFM nearest-neighbour exchange $J_A$ between the zig-zags. In the dimer model, the nearest-neighbour FM interactions within a zig-zag chain are replaced by a stronger intra-dimer FM interaction $J_{FS}$ and a weaker inter-dimer FM interaction $J_{FW}$. Between the zig-zag chains are additional inter-dimer interactions denoted $J_{A1}$ and $J_{A2}$ as they link antiparallel spins. The symmetry of the model does not require them to be identical, nor is it necessary for both of them to be AFM. Finally, in both models there is an inter-layer interaction coupling the two sheets of the bilayer along the crystallographic $c$ direction, denoted $J_{\perp}$. The magnetic coupling between bilayers is negligible; consequently the spin wave spectrum is dispersive only in the $a-b$ plane and so can be treated as quasi-2D.

\begin{table} \renewcommand{\arraystretch}{1.3}
\begin{center}
\small{  
  \setlength\tabcolsep{1pt}
  \begin{tabular}{@{\extracolsep{\fill}}rrc|rrc}
  \hline
      \multicolumn{3}{c|}{Goodenough model}    & \multicolumn{3}{c}{dimer model}            \\
                 & Ref (meV)    & Range (meV)  &              & Ref (meV)     & Range (meV) \\
  \hline
      $J_{F1}$   &   -11.39(5)  &  [-20, 0]    &   $J_{FS}$   &    -14.20(8)  & [-20, 0]    \\
      $J_{A}$    &     1.50(2)  &  [0, 3]      &   $J_{FW}$   &     -8.43(6)  & [-20, 0]    \\
      $J_{F2}$   &    -1.35(7)  &  [-3, 3]     &   $J_{A1}$   &      1.52(1)  & [-3, 3]     \\
      $J_{F3}$   &     1.50(5)  &  [-3, 3]     &   $J_{A2}$   &      1.52(1)  & [-3, 3]     \\
   $J_{\perp}$   &     0.88(3)  &  [0, 3]      & $J_{\perp}$  &      0.92(3)  & [-20, 0]    \\
      $D$        &     0.074(1) &  [0, 0.2]    &   $D$        &      0.073(1) & [0, 0.2]    \\
  \hline
  \end{tabular}
}
  \caption{
Spin wave exchange parameters (meV) for the Goodenough and dimer models for Pr(Ca$_{0.9}$Sr$_{0.1}$)$_2$Mn$_2$O$_7$ with standard errors in parentheses as determined by~\cite{johnstone2012}, and the range of values used to generate random datasets for training the neural network classifier in this work. Positive values indicate antiferromagnetic exchange. $D$ denotes the single-ion anisotropy term, with positive values here indicating an easy-plane anisotropy.
}
  \label{tab:tab1}
\end{center}
\end{table}

\section{Data}
\label{sec:data}

\subsection{Experimental data} \label{sec:exp}

The experimental INS data was acquired using the MAPS spectrometer at the ISIS Neutron and Muon Source, and the conventional analysis of that data was previously published~\cite{johnstone2012}. Datasets were collected with several different incident neutron energies and monochromating (Fermi) chopper speeds ($E_i=25$, 35, 50, 70, 100, 140~meV, with corresponding speeds $f=300$, 200, 200, 250, 300, 400~Hz) to span the full bandwidth of the spin wave spectrum in a series of energy transfer windows with appropriate wave vector and energy resolution. The co-aligned array of single crystals of Pr(Ca$_{0.9}$Sr$_{0.1}$)$_2$Mn$_2$O$_7$ was mounted with the incident beam parallel to the $c$ axis so that the $a-b$ plane is imaged in the detectors. All measurements were carried out in a closed cycle refrigerator at 4~K. For each dataset, an estimate of non-magnetic scattering and background was made from several cuts as a function of energy transfer ($\hbar\omega$) at different wave vectors, from which sections were stitched together to construct a single cut that included no spin wave scattering.

\begin{figure*}
\centering
    \subfigure{\includegraphics[width=0.8\textwidth,viewport=8 7 855 431]{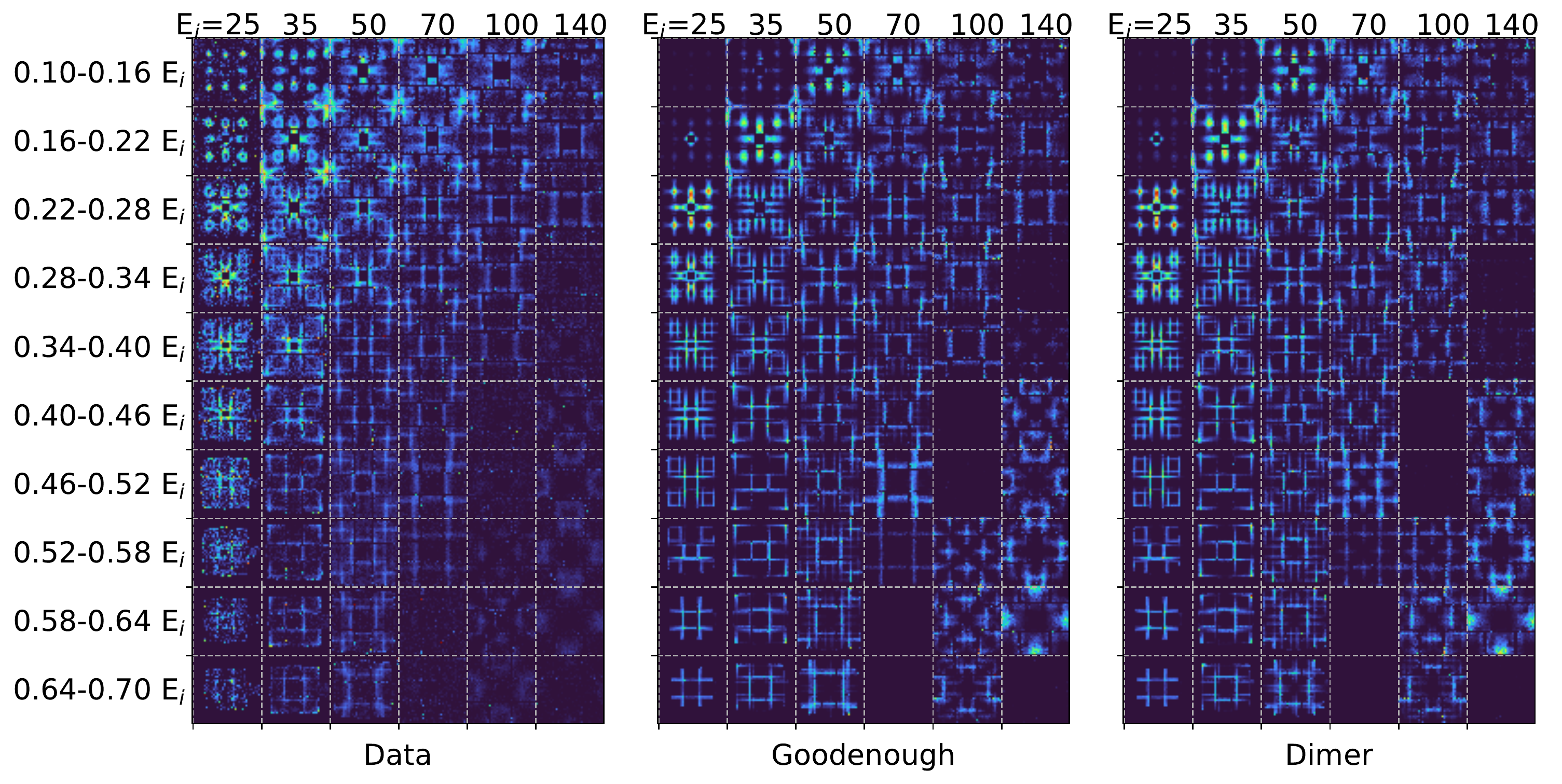}}
\caption{\label{fig:experimental} 2D representation of the experimental data (left panel). The data are arranged column-wise in terms of incident neutron energy ($E_i$), with $E_i$ in meV given at the top of the column. The data are arranged row-wise into bins of energy transfer $\hbar\omega$=0.10-0.16$E_i$, 0.16-0.22$E_i$ etc. The middle and right panels show the calculated spectra using the Goodenough (middle) and dimer (right) exchange models at the published parameters noted in Table~\ref{tab:tab1}.}
\end{figure*}

To use with the convolutional neural networks described below, the background-subtracted data were arranged into a gray scale image, obtained by first vertically stacking ten constant-energy transfer slices per incident energy spectrum, each stack uniformly dividing the energy transfer range between 0.1$E_i$ and 0.7$E_i$ into slices with thickness 0.06$E_i$. We chose to exclude data close to the elastic line (i.e. zero energy transfer) as the signal there is dominated by non-magnetic scattering, and also to exclude the spectrum above 0.7$E_i$ as the signal there has a significant non-magnetic background. The constant-energy transfer slices are gray scale 40$\times$40 pixel images over the wave vector range $-1 < h, k < 1$, which covers the full periodicity of the spin waves in the reciprocal lattice. The ten slices for each of the six $E_i$ form 40$\times$400 pixel strips, which were then stacked horizontally producing a 240$\times$400 pixel image, shown in Figure~\ref{fig:experimental}. This image was used as input to the trained neural network classifier in order to determine which exchange model (Goodenough or dimer) better describes the measured data.

\subsection{Training Data}

The convolutional neural networks described in the next section were trained on simulated data, in the same 240$\times$400 gray scale image format as the experimental data. For the training datasets we used linear spin wave theory, as implemented in the \emph{SpinW} code~\cite{toth_spinw}, to simulate the spin wave spectrum corresponding to sets of distinct spin Hamiltonian parameters $\{J_{ij}, D\}$. Each constant energy transfer slice (40$\times$40 pixel sub-image) is computed independently by convolving the spin wave theory spectrum with the spectrometer resolution function for each detector element - energy transfer bin in the data, using a Monte Carlo integration method~\cite{tobyphd} (named ``Tobyfit" after the first program to implement it), as implemented in the neutron scattering data analysis code \emph{Horace}~\cite{horacepaper}. The resulting simulation is sliced to produce a stack of constant energy transfer slices in precisely the same way as the experimental data are sliced. This resolution function calculation properly accounts for the functional form of the contributions from individual instrument components. However, the resolution convolution is quite computationally intensive, requiring approximately 12~CPU-hr per parameter set (240$\times$400 image). The largest amount of time is spent by the spin wave calculation itself, that is, within \emph{SpinW}, because the resolution convolution requires the calculated structure factor at a large number of wave vector and energy transfer $(\mathbf{Q}, \hbar\omega)$ positions, each of which requires the computation and diagonalisation of a Hamiltonian matrix and additional matrix multiplications to obtain the spin-spin correlation function and hence neutron scattering intensity. In order to speed this up we used another application, \emph{Brille}~\cite{brille}, to pre-compute these structure factors in a dense grid within the first Brillouin zone, and then to linearly interpolate within this grid for the required $(\mathbf{Q}, \hbar\omega)$ points needed by the resolution convolution. This reduced the calculation time to $\approx$1~CPU-hr per parameter set.

We also explored an approximate and faster resolution convolution method, motivated by the general desire to reduce the number of CPU hours needed when using ML techniques. Instead of using expensive Monte Carlo integration, the resolution function covariance matrix is pre-computed for each incident energy spectrum in a grid in $(\mathbf{Q}, \hbar\omega)$ space and then used to define a 4D Gaussian profile with which the spin wave theory dispersion is convolved. With an 80$\times$80 grid per 40$\times$40 pixel slice the calculation times were reduced to $\approx$10~CPU-min per parameter set. However, the resulting calculated spectra \rem{did not match well with the data and }yielded larger classification errors as described below. The coarseness of the grid resulted in an underestimate of the resolution broadening, and consequently the calculation is ``sharper" than the data and the energy transfer of the top (bottom) of the spin wave dispersion bands were calculated to be too low (high) in energy transfer. The problem can be rectified by using finer grids but this then results in similar calculation times to the ``Tobyfit" method with \emph{Brille} interpolation. Furthermore, the use of a pre-computed grid also results in aliasing effects for certain grid sizes which the Monte Carlo method is not susceptible to. The important lesson from this experience is that accurate account of the resolution function needs to be taken for robust interpretation of INS, despite the concomitant cost in computing resources.


We generated 3322 images for each of the Goodenough and dimer models, split 6000:644 between training and validation datasets, which required a total of $\sim$7000~CPU-hr. The parameters for these datasets are randomly generated by independently selecting from a uniform distribution within the limits noted in Table~\ref{tab:tab1} for each $J_{ij}$ and model. These limits bracket likely maximum ranges of estimates of the parameters from the bandwidths of the spin wave branches and values of the exchange constants determined in various other manganites. This mimics the procedure that would be followed if the training datasets were being generated for an unknown system.

The data can be downloaded from an online repository~\cite{datarepo}. Figure~\ref{fig:experimental} shows the calculated spectra using the Monte Carlo resolution convolution method for both the Goodenough and dimer models alongside the measured data.

\section{A network for phase discrimination}
\label{sec:nn}
\subsection{Can a neural network learn to distinguish phases?}
\label{sec:sub-nn}

We first want to establish that we can train a neural network (NN) capable of learning to distinguish between the Goodenough and dimer models based on the INS spectrum. This is a classic inverse problem; while the forward mapping from the spin Hamiltonian for each model to the INS spectrum is well known, the inverse is not.


Analysing the spin wave spectrum pixel-by-pixel using a multi-layer perceptron (MLP) type NN architecture would lead to a rapidly exploding number of parameters. 
In addition, as an MLP treats the input image as a vector of data, it can be very sensitive to small shifts in spectral feature positions which may not be characteristic of the different exchange models. 
While these shifts do not pose a problem for the training of the network which is done on generated data, it may become an issue when confronted by experimental data, because slight miscalibrations of the spectrometer may produce data which is offset compared to the ideal  simulations.
Furthermore, due instrumental resolution broadening, neighbouring pixels in a spectrum (image) are correlated, and a MLP would have to learn these correlations explicitly. Instead, 
we employ the popular convolutional neural network (CNN) architecture \cite{lecun1989backpropagation, fukushima1982neocognitron}, whereby the image (spectrum) is first passed through a series of filters, before feeding the values of these filters into a MLP architecture. During training the weights of the filters and the MLP are updated. The filters can be seen as learning a reduced representation of the spectrum (including implicit correlations between neighbouring pixels), 
while the MLP learns the function to map these features to the corresponding exchange interaction model.

\begin{figure}
\begin{center}
\includegraphics[width=\columnwidth]{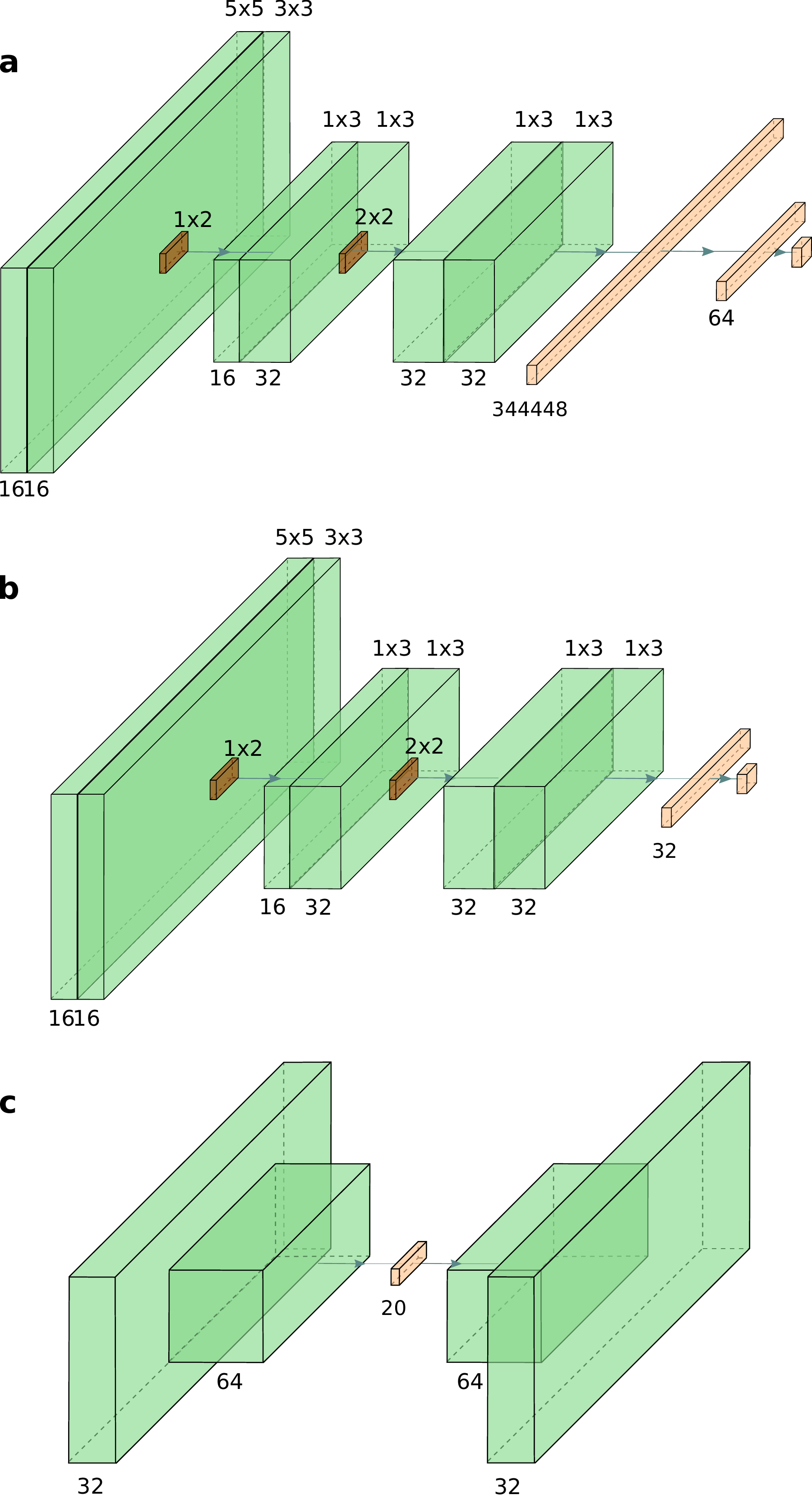}
\end{center}
\caption{Network architectures. (a) The convolutional neural network (CNN) architecture; at each (green) convolution layer the number of filters is indicated below and the size of the convolution matrix above the blocks, and at each fully connected (orange) layer the number of input dimensions is indicated. After each block of convolution layers we perform a MaxPool (brown block with pool size indicated) and batch normalisation (see text). The large dense layer (344448 element orange block) is a vector obtained by flattening the outputs of the final layer of convolution filters, this layer is then connected to a 64 node layer and finally to the classification head. (b) The same CNN as (a), but rather than flatten the final layer of filters a global pooling of each of the 32 filters is used to generate 32 values which are connected directly to the classification head; this architecture is used to construct class activation maps. (c) The variational autoencoder architecture; at each convolution (green) layer the number of filters is indicated, we also perform $(2 \times 2)$ striding (the convolution filter skips across the input in steps of two, thereby halving the size of the output) at each convolutional layer, the central latent (orange) layer consists of $20$ Gaussian distributions as characterised by a mean and standard deviation.\label{fig:arch}}
\end{figure}

There are a wide array of CNN architectures already trained and available for image recognition problems, for example the \alexnet\cite{krizhevsky2012imagenet} and \resnext \cite{xie2017aggregated} architectures which are both previous winners of the \imagenet \cite{deng2009imagenet} data classification challenge. In this work we build a relatively simple CNN architecture, as the data that we are classifying and the number of potential classes are quite small. As we will demonstrate later, choosing simpler architectures can help with interpretation of the final CNN inference.

The architecture we built is represented schematically in Figure~\ref{fig:arch}~(a). The convolution layers are coupled with a pooling layer so that the image is subsequently down-sized as it passes through the convolutions and a batch normalisation is applied after pooling. In batch normalisation the outputs from a layer are normalised to fall between 0 and 1 across a pre-defined batch size of training samples. In MaxPool the size of a filter is reduced between layers by pooling sections of the filter and taking the maximum value of each pool. At the final convolutional layer the values of each filter $\mathcal{F}^j_{xy}$ (where $x$ and $y$ are spatial coordinates and $j$ is the filter index) 
are flattened to produce a $117\times92\times32=344448$-element vector~\footnote{A convolutional filter with size $n$, padding $p$ and stride $s$ changes an image with input dimension $d$ to $(d-n+2p)/s + 1$; in our classification CNNs $p=0$ and $s=1$ for all cases. That is, after the first $5\times 5$ convolutional layer, the image (initially $240 \times 400$) size is $236\times 396$ and after the first $1\times 2$ MaxPool its size is $234\times 197$. The number of filters we used at each layer is specified in Figure~\ref{fig:arch}. The final convolution layer has 32 filters and the image has been downsized to $117 \times 92$ pixels.} which is fed into a fully-connected layer (MLP) with $d=64$ output nodes. 
Classification is then performed by a weighted summation of these 64 values $v_i$ based on trainable weights $w^c_i$ and a trainable bias $b_c$, which serve as input to a sigmoid function $\phi$. 
That is, the score for class $c$ is
\begin{eqnarray}
Y^c = \phi(\sum_{i=1}^{i=d} w^c_i v_i + b_c)
\label{eqn:eqn1}
\end{eqnarray}
\noindent with the class with the highest score selected by the network.

The simulated training data are split into training and validation sets as described in the data generation section. 
To save memory the training dataset is divided into batches of 32 images. In a single training iteration, the network is fed these 32 images to calculate the loss function (we use the mean binary cross-entropy loss of the classification of all images in the batch as the loss function). The gradient of each adjustable weight in the network with respect to the loss is then calculated using backpropagation and the weight is adjusted in the direction of steepest descent. This iteration is repeated for each batch of 32 images, until all batches have been passed through the network once, which is termed an epoch. At the end of each epoch the loss from the separate validation dataset is calculated. If this validation loss is much larger than the loss calculated from the training set after adjusting the weights, the network is overfitting the training data such that it cannot generalise its inference to the similar data in the validation set.
The training is set to terminate if the validation accuracy has not improved for 20 epochs, up to a maximum of 500 epochs.
The full details of the network training and the code for the architecture are available in an online repository~\cite{datarepo}. The final trained CNN achieves $> 96\%$ accuracy in predicting the correct exchange model for data in the validation set for both Goodenough and dimer datasets. 
The training curves are included as supporting information.

\subsection{How much can we trust the predictions?}
\label{sec:duq}

When a network has been trained to separate data into two classes, it can be difficult to know how reliable a prediction on a new piece of data is. Binary classification (Equation~\ref{eqn:eqn1}) applies a sigmoid function, which intentionally favours output values close to either 0 or 1, forcing separation between classes; however, this means that an individual sample may be classified strongly into a class despite a degree of ambiguity in the result. 
For example, when we trained the previously described network on data simulated with the Monte Carlo resolution convolution, and then feed it the experimental data spectrum, it gives class scores corresponding to  $[Y^{\mathrm{dimer}}, Y^{\mathrm{Goodenough}}]$ of $[0.012, 0.988]$. However, when the same network was trained using data simulated using the faster pre-computed grid of covariance matrix resolution convolution method, it gives $[0.999, 0.001]$. That is, it not only wrongly classifies the experimental data as being in the dimer class, it does so strongly.
It is thus 
incorrect and misleading to interpret this output vector of the sigmoid classification as a measure of confidence for that classification. To address this concern we have adapted the recently published deterministic uncertainty quantification (DUQ) scheme to work with our classification network~\cite{van2020uncertainty}.

There are a range of methods available for uncertainty quantification in NN predictions, for example Bayesian neural nets (BNNs) \cite{neal2012bayesian}. Pure BNNs are intractable for exact inference, although a number of approximations have been proposed, including the simple-to-implement MC Dropout approach \cite{gal2016dropout}. In practice most of these Bayesian approaches are outperformed by the Deep Ensembles (DE) approach \cite{lakshminarayanan2017simple}, in which multiple networks with the same architecture are trained from different initial values of weights and biases and dataset orderings, resulting in an distribution of answers. The drawback of the DE approach is the requirement to train and run multiple networks and the associated linear growth of computational cost. We instead choose the DUQ approach, which can provide uncertainty estimates in a single forward pass of the network, but shows performance in out-of-distribution data detection (identifying cases where predictions would be an extrapolation from training data, rather than an interpolation) on a par with DE approaches \cite{van2020uncertainty}. We note that several new approaches show good potential for uncertainty quantification in NNs, such as the stochastic differential equation approach \cite{kong2020sde} and radial BNNs \cite{farquhar2020radial}, but a full comparison of all approaches is beyond the scope of the current study.

\begin{figure}
\includegraphics[width=\columnwidth]{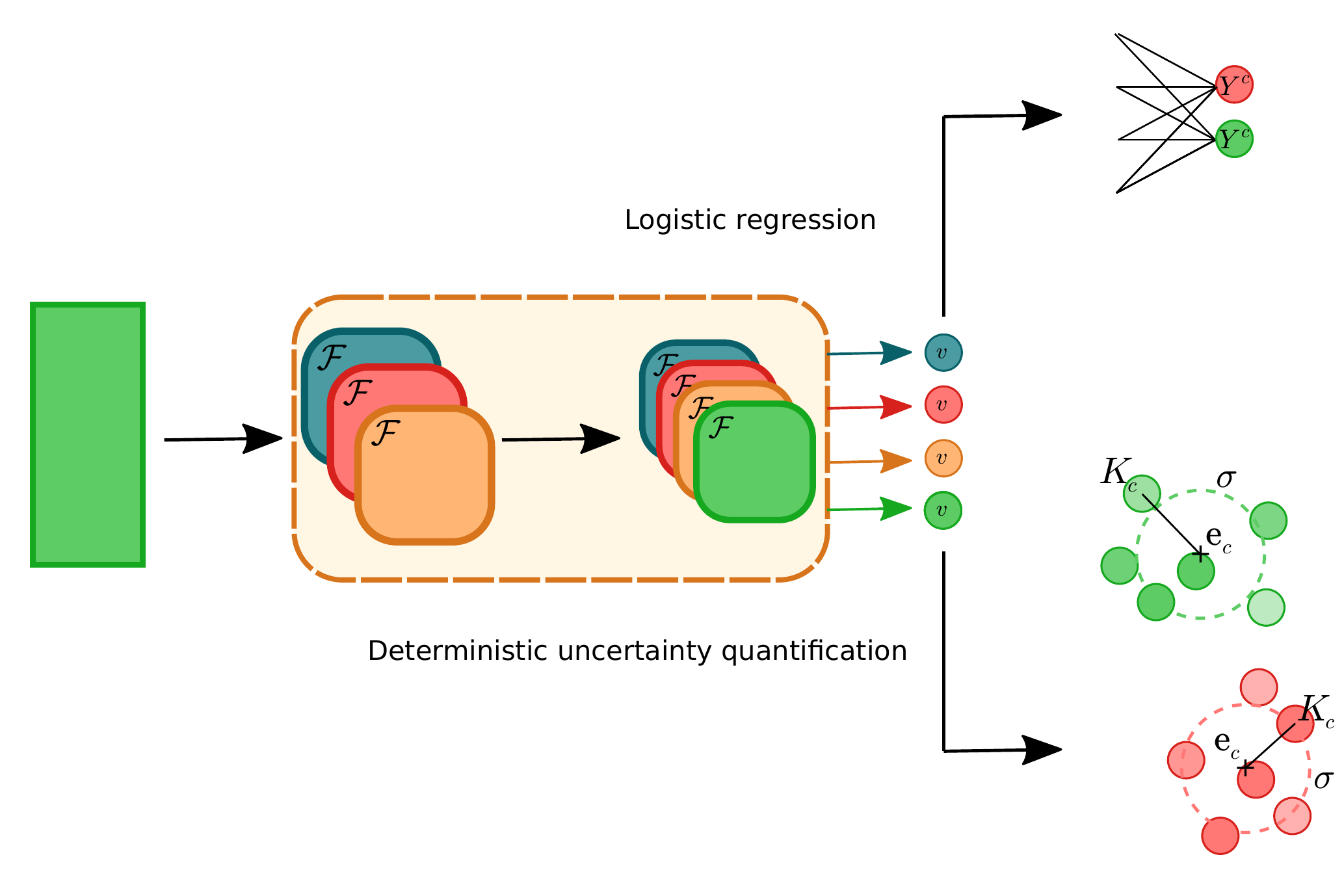}
\caption{Schematic representation of the deterministic uncertainty quantification (DUQ) method. The input initially passes through a series of convolutions (orange block) to extract features. In standard logistic regression the outputs from the convolutions are classified by summing the weights connecting each filter $f_i$ to the class $\mathcal{C}$ of interest, as in equation~\ref{eqn:eqn1}. DUQ clusters the examples based on distances $K_c$ in a high dimensional space of the outputs from the convolutions from the centre $\mathbf{e}_c$ of clusters of training examples, according to equation~\ref{eqn:eqn2}. \label{fig:duq}}
\end{figure}


The principle of DUQ as applied here is as follows. Let $M_{\Theta}: \mathbf{x}\in\mathbb{R}^n \rightarrow \mathbf{v}\in\mathbb{R}^d$ be the transformation learnt by the network which maps an input vector $\mathbf{x}$ (the input image, size $240\times400$ pixels) with dimension $n=96000$ to an output features vector $\mathbf{v} = \{v_1, \ldots, v_d\}$ with dimension $d=64$ (the number of output nodes connected to the classification head), and $\Theta$ are the parameters (weights and filters) of the network.
In DUQ~\cite{van2020uncertainty} (Figure~\ref{fig:duq}), instead of the final dense layer (the classification head) which maps $\mathbf{v}$ to the class scores $\mathbf{Y} = \{Y^1, \ldots, Y^c\}$, a different mapping is learnt which transforms $\mathbf{v}$ into a vector space with dimension $m$ where the samples of the same class $c$ are clustered together.
This is embodied by a \emph{per-class} learnable weight matrix $\mathbf{W}_c \in \mathbb{R}^{m \times d}$ with $m=64$ in our case. The correlation $K_c$ between this $\mathbf{W}_c \mathbf{v}$ vector and the class centroid $\mathbf{e}_c$, given by


\begin{equation}
    K_c(\mathbf{v}, \mathbf{e}_c) = \exp \left[-\frac{||\mathbf{W}_c \mathbf{v} - \mathbf{e}_c||^2_2}{2n\sigma^2}\right] \ ,
\label{eqn:eqn2}
\end{equation}

\noindent is then a measure of the uncertainty in the classification, with the hyper-parameter $\sigma$ being a characteristic length scale, also learned during the training. The network then assigns a given input sample $\mathbf{x}$ to the class with the largest correlation $K_c$.
Thus instead of classifying the sample based on the highest score $Y^c$ from equation~\ref{eqn:eqn1}, it is classified as belonging to the centroid closest to it, with a confidence $K_c$ that corresponds to the Euclidean distance between the new point and the centroid. In this way, data that are out of the training distribution are far from any of the trained centres of gravity and the confidence for classification is low. 

The network architecture we now use to classify the experimental neutron scattering data is the same one demonstrated in the previous section to reliably predict the correct exchange model with the validation datasets, and is depicted in Fig~\ref{fig:arch}~(a). The only difference is that the vector at the 64 node layer is then used as the input to the DUQ analysis. This CNN was trained for 100 epochs on data prepared as described in the Training Data section.

Networks were trained on the \rem{two different }simulated datasets for both the Goodenough and dimer models, generated by 
the resolution convolution methods described earlier: one using the fast but approximate method using a pre-computed grid of covariance matrices ($M^{\mathrm{pre}}_{\Theta}$), and the other using the more expensive but accurate Monte Carlo integration ($M^{\mathrm{MC}}_{\Theta}$) method. We then use these networks to classify the background-subtracted experimental spectrum. The DUQ networks yields a correlation value between 0 and 1 to indicate the distance in hyperspace between the weight vector associated with the input and the output classes. 

The network trained on pre-computed resolution data $M^{\mathrm{pre}}_{\Theta}$ gives an output corresponding to $[K_{\mathrm{dimer}}, K_{\mathrm{Goodenough}}]$ of $[0.73, 0.49]$ , showing that the network correlates this input more with the dimer model than with the Goodenough model. This classification is incorrect; moreover, the correlation values suggest the the network does not distinguish between the two models with great certainty and the experimental data lie out of distribution of the training data.

\begin{figure}
\includegraphics[width=\columnwidth]{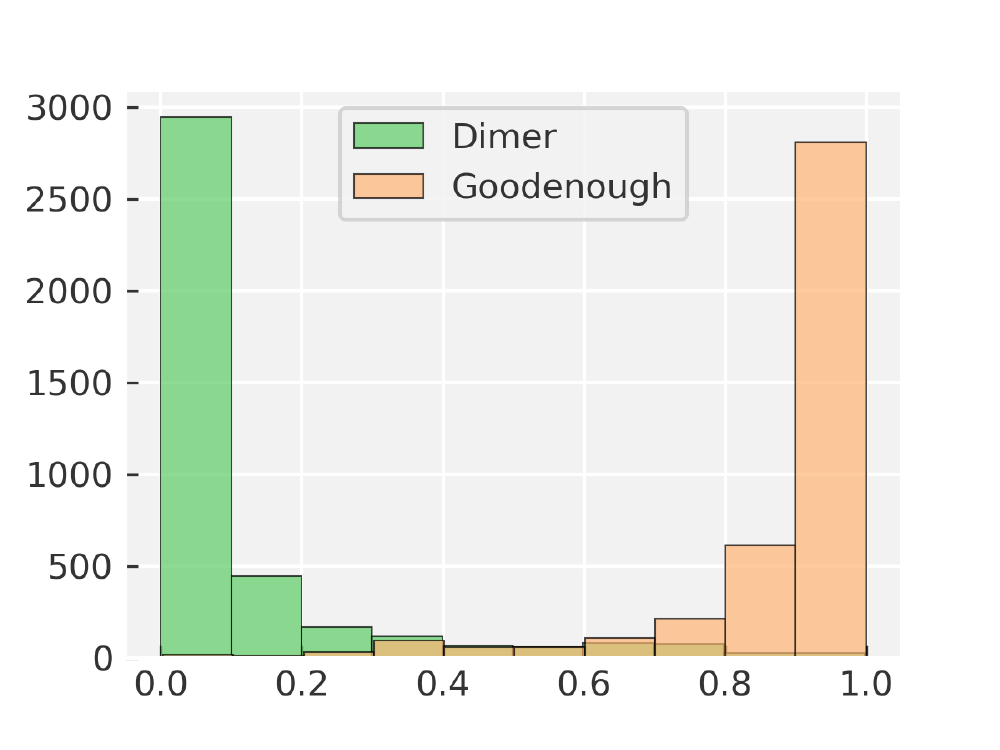}
\caption{The correlation values $K_c$ for the $c=\mathrm{Goodenough}$ category assigned by the 
$M^{\mathrm{pre}}_{\Theta}$
DUQ classifier.
The plot is a histogram of values of the Goodenough class output given by the neural network on the full training datasets of spectra generated using the Goodenough (orange) and dimer (green) models (3322 spectra each).
\label{fig:correlation}}
\end{figure}

To demonstrate this point further we have calculated the correlation values $K_{\mathrm{Goodenough}}$ that the network assigns to the Goodenough category for every spectrum in the training set for $M^{\mathrm{pre}}_{\Theta}$. The correlation values are presented in Figure~\ref{fig:correlation}, demonstrating clearly that on data corresponding to the training distribution the network will overwhelmingly predict values very close to 1.0 for the (ground truth) Goodenough class and close to 0.0 for the dimer class. This further emphasises that the result from $M^{\mathrm{pre}}_{\Theta}$ on the background subtracted experimental data indicates that these data are not within the training distribution.

In contrast, when the network trained on Monte Carlo resolution-convolved data $M^{\mathrm{MC}}_{\Theta}$ was tested with the background subtracted experimental spectrum, the output corresponding to the classes $[K_{\mathrm{dimer}}, K_{\mathrm{Goodenough}}]$ was $[0.05, 0.99]$. The $M^{\mathrm{MC}}_{\Theta}$ network not only predicts the correct result, but does so with a high degree of confidence, indicating that the training data with the Monte Carlo resolution convolution method covers a distribution encapsulating the experimental data with the background subtracted.

This result highlights two important points. First the inclusion of accurate representations of instrument and experiment resolutions is critical if one wishes to train a neural network on simulated data and later to apply it to analyse experimental results. Second having a method to quantify the confidence of the machine learning is extremely important. In a standard CNN with logistic regression rather than DUQ the $M^{\mathrm{pre}}_{\Theta}$ would have classified the spectrum as belonging to the dimer class, with no indication that there was uncertainty in this classification or that the data were out of distribution.

To further demonstrate the utility of the DUQ method for identifying out of distribution data we have tested the $M^{\mathrm{MC}}_{\Theta}$ network on some ``Franken-data''. The Franken-data are generated by stitching together spectra from several different materials measured with the same instrument settings (and hence instrument resolution) as the PCSMO data. The resulting image, constructed to have the same dimensions as the PCSMO data, looks like a reasonable INS spectrum at first glance. The $M^{\mathrm{MC}}_{\Theta}$ network gives correlation values of $[0.35, 0.85]$, where the closeness of the two values is indicative of a sample outside of the training distribution, indicating the utility of DUQ for identifying cases where the classification should not be trusted.

\subsection{Why does the network predict what it does?}
\label{sec:cam}

We have established that a CNN is capable of performing the inverse problem of inferring the magnetic exchange model from the simulated spectrum. However, a highly pertinent question is: can the network tell us what influenced this decision? If we can address this question we can gain confidence in the network, and can also use it in a predictive fashion, so that in advance of analysing data the CNN could be used to work out what are the most important regions for attention in the data. Moreover, knowing what in the data influenced the decision is what helps yields scientific insight.

\begin{figure}
\includegraphics[width=\columnwidth]{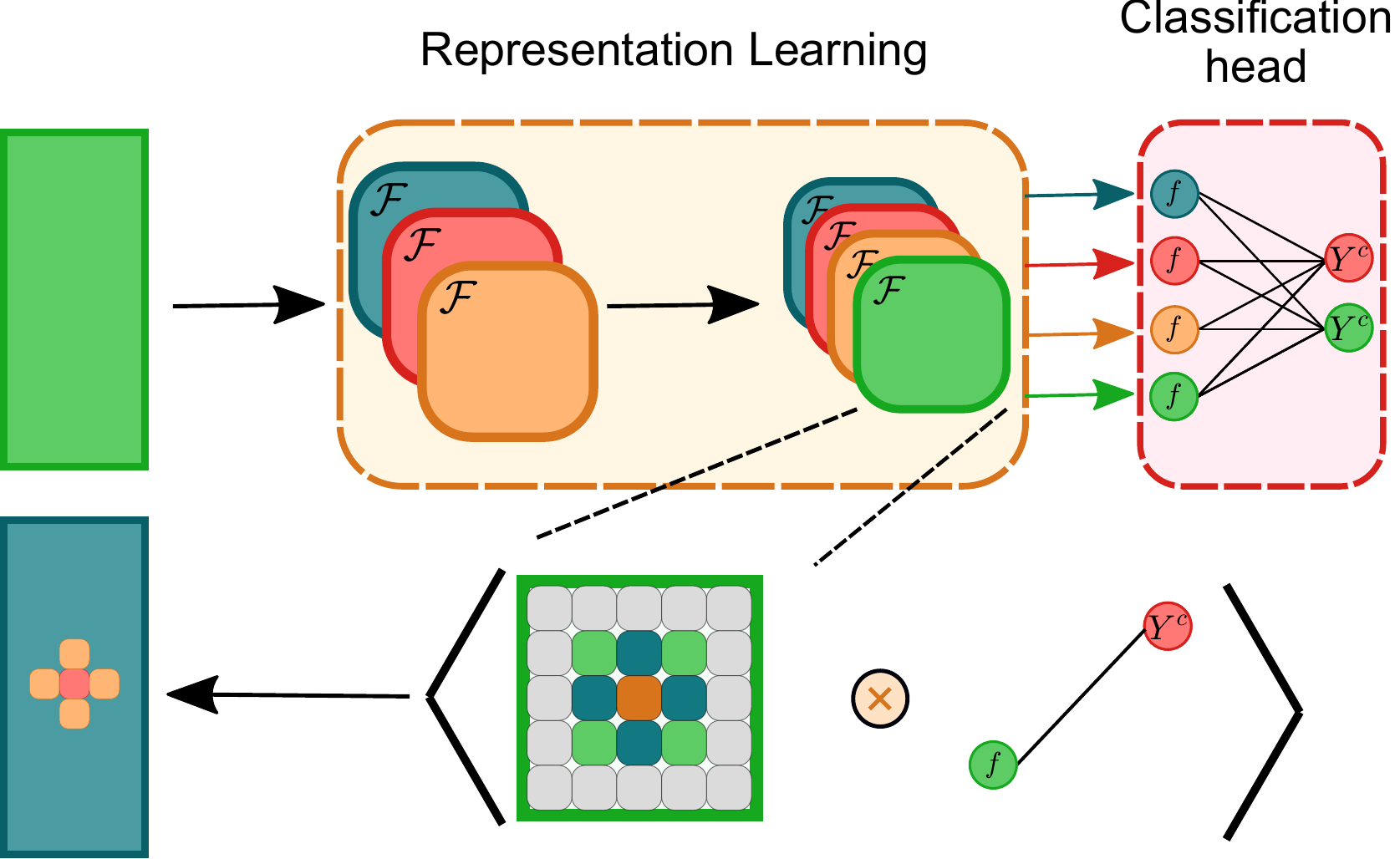}
\caption{\label{fig:fig2} Class activation map: The input initially passes through a series of convolutions (orange block) to extract features. The final layer of filters ($\mathcal{F}_{xy}^j$ summed to give $f^j$) are fed into the classification head as the global average of each filter to do classification (red block). To build an activation map, we take the cross product of the filter weights with the activation linking the average of that filter to the class of interest. For a single image we take the average of all of these cross products to build a map of why that particular class was activated, equation~\ref{eqn:eqn4}.}
\end{figure}

To ask the network which are the important regions for making a distinction we employ class activation maps (CAMs)~\cite{Zhou_2016_CVPR, selvaraju2017grad}. CAMs relate regions of an input signal to the output classification. The mechanism for producing a CAM is outlined in Figure ~\ref{fig:fig2}.
In this example, the final convolutional layer is directly connected to the classification head (a dense fully-connected layer), so the extent to which a filter $\mathcal{F}^j_{xy}$ is responsible for a classification $Y^{c}$ is determined by the weight $w_j^c$ connecting the $j^{\mathrm{th}}$ filter to the $c^{\mathrm{th}}$ class. The activation map is then obtained by:
\begin{equation}
    A^c_{xy} = \frac{1}{n_f}\sum_{j=1}^{j=n_f}w_j^c \mathcal{F}^j_{xy} \ .
\end{equation}
That is, the image $A^c_{xy}$ is the average (denoted $\langle \ldots \rangle$ in Figure~\ref{fig:fig2}) of the $n_f$ filters weighted by their contributions to the classification $c$.

This approach however, cannot work with the classification architecture we introduced in section~\ref{sec:sub-nn} where there is an intervening fully connected layer between the final convolutional layer and the classification head, because it requires the weights $w_j^c$ implying that the $j^{\mathrm{th}}$ filter directly impact the class $c$. Thus, we use the Grad-CAM~\cite{selvaraju2017grad} method where instead of the weights $w_j^c$, the gradients $\frac{d Y^c}{d f^j}$ are used where $f^j=\sum_{xy}\mathcal{F}^j_{xy}$ is the sum of all pixels (global pool) of a filter. These gradients can be calculated through any number of intermediate layers between the last convolutional layer and the classification head using backpropagation. Thus, the activation map we used is given by:

\begin{equation}
A^c_{x,y} = \frac{1}{n_f} \sum_{j=1}^{j=n_f} \frac{d Y^c}{d f^j} \mathcal{F}^j_{xy} \ .
\label{eqn:eqn4}
\end{equation}

Applying this Grad-CAM to the original architecture, depicted in Figure~\ref{fig:arch}~(a), yielded a class activation map which 
does not reveal any activation patterns which correspond to physical intuition. 
Thus, whilst the extra complexity from the intermediate densely connected layer between the final convolution layer and the classification head allows the classification network to be flexible and to produce an accurate determination between the Goodenough and dimer models, 
it also obscures the relationships between the spatial inputs of the model and the outputs. 

We next build a simpler network architecture (Figure~\ref{fig:arch}~(b)), in this case rather than flattening the final convolutional layer we take the global average of each of the filters and make an input vector of size 32 (one for each filter) and connect this directly to the classification head of the network. This means that the filters now directly contribute to the classification and the more activated a filter is, the more it contributes. However, when presented with the experimental spectrum this network classifies with very little certainty, as the new network is less flexible and less robust to experimental noise.

\begin{figure}
\centering
\begin{subfigure}
    \centering
    \includegraphics[width=\columnwidth]{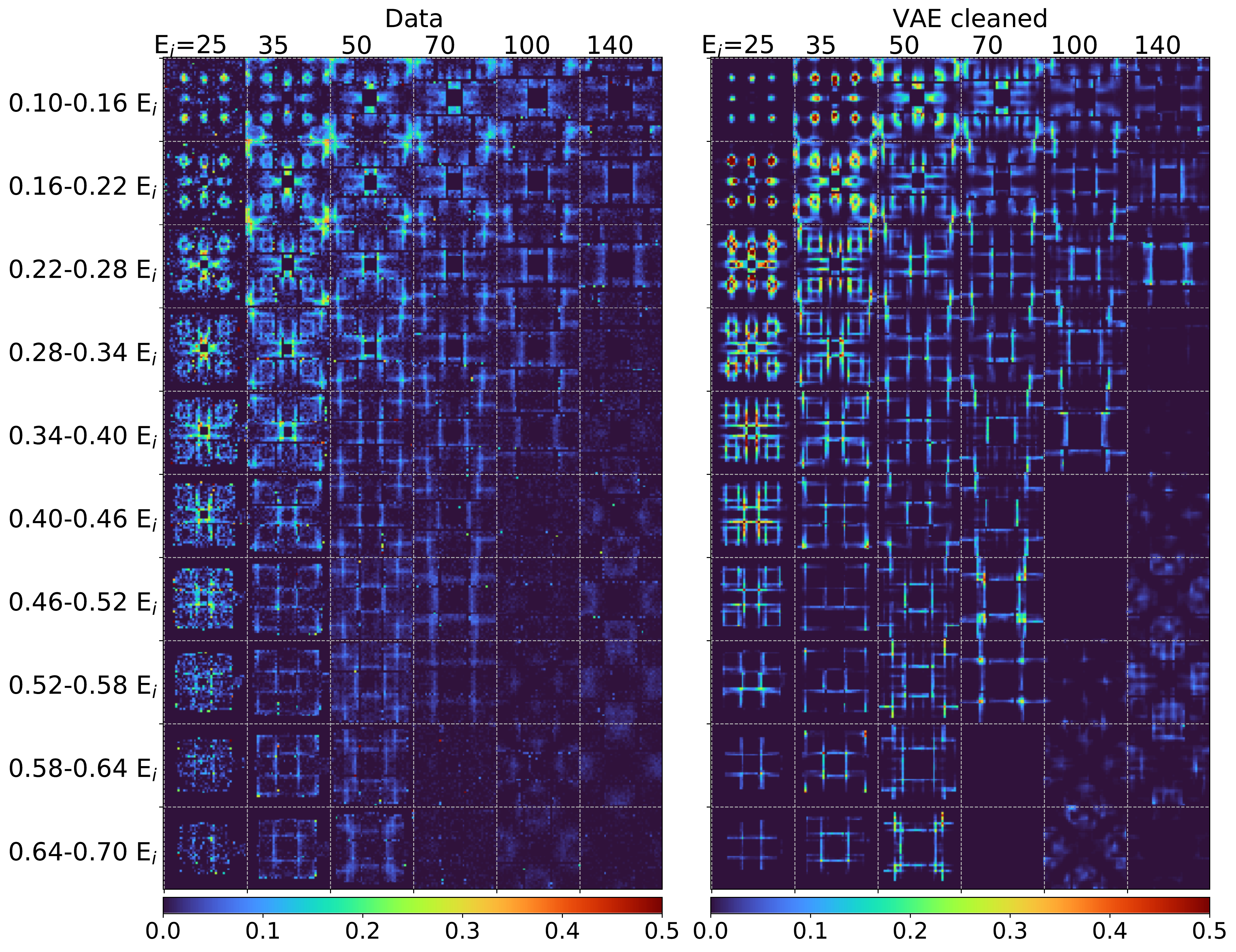}
\end{subfigure}
\begin{subfigure}
    \centering
    \includegraphics[width=\columnwidth]{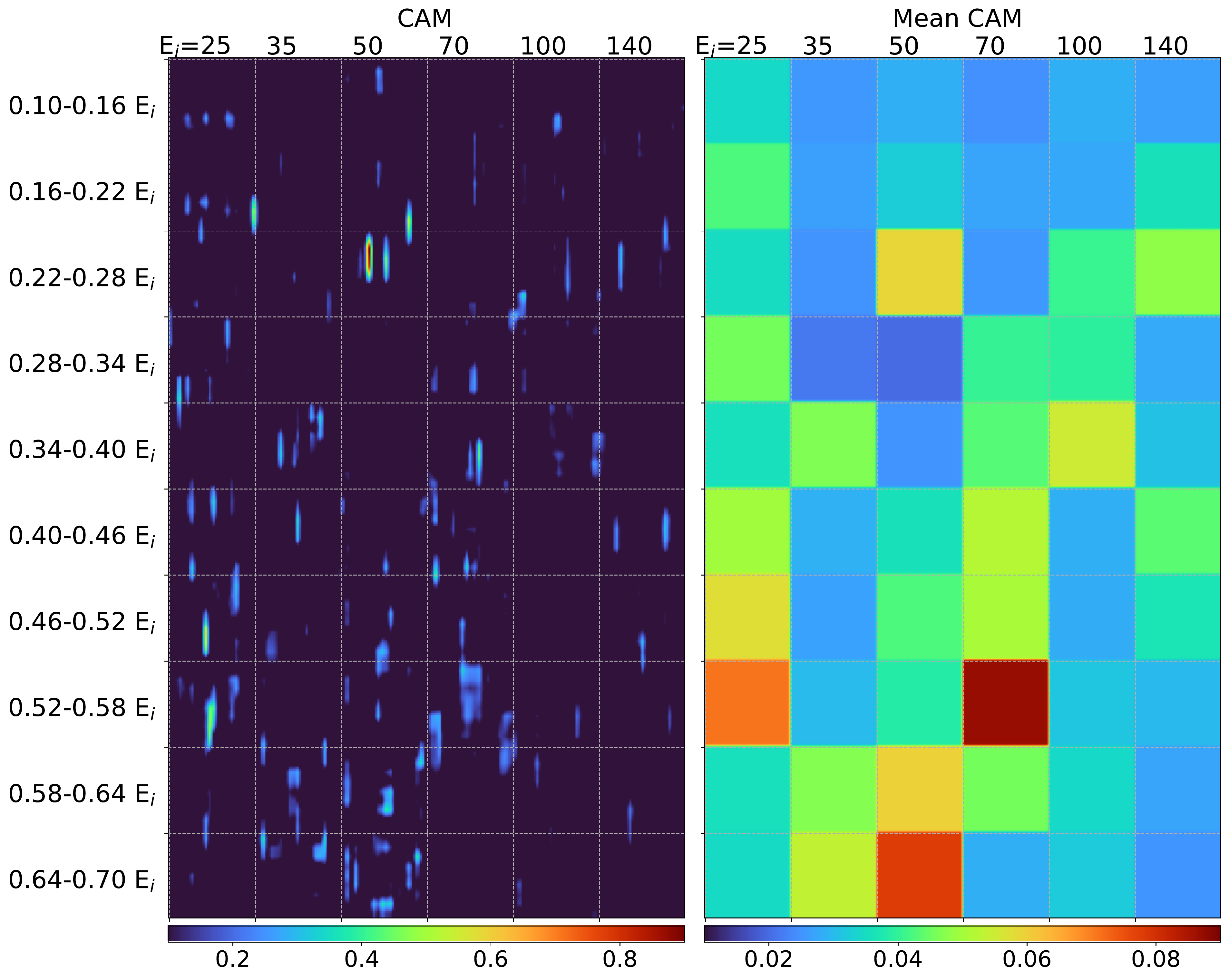}
\end{subfigure}
\caption{\label{fig:fig3} Upper: the effect of cleaning the experimental spectrum with a variational autoencoder, left is the experimental spectrum, right is the same spectrum passed through the trained autoencoder. Lower: Class activation maps. Left shows the class activation map for the autoencoder cleaned signal, right we have taken the mean of the activations per slice of the spectrum and plotted these to show which slices contributed most strongly to the classification of Goodenough.}
\end{figure}

To overcome the sensitivity to experimental noise we build a variational autoencoder (VAE) to clean the signal~\cite{kingma2013auto}. Previous work has shown the application of autoencoders for cleaning diffuse (i.e. energy integrated) neutron scattering signals~\cite{samarakoon2020}. We build a VAE with two convolutional layers in the encoder, with strided convolutions in order to compress the image size - the architecture is shown in Figure~\ref{fig:arch}~(c). At the ``bottle-neck'' or ``latent'' layer we have 20 units. The network is then trained on 3500 examples of simulated dimer and Goodenough datasets (1750 of each) with Monte Carlo resolution convolution, the training is run for 100 epochs to maximise the evidence lower bound (ELBO) loss\cite{kingma2013auto, lin2019balancing}. For a VAE the network trains to reproduce the input, so no labels are required. ELBO balances a loss on the reconstruction (how similar output is to input) with the divergence between the inferred distribution in the latent space and the prior distribution (normal distribution with mean 0 and variance 1) of the latent space. Because the data pass through the reduced dimensionality of the latent layer the information content is necessarily reduced and only the stronger features of the initial signal are retained. In this way noise can be removed from the original, the result of cleaning the signal with the autoencoder is shown in Figure~\ref{fig:fig3}. The full neural network architecture and training routine are available in the repository associated with this paper~\cite{datarepo}.

Figure~\ref{fig:fig3} shows the experimental spectrum, with the background subtracted (top left) and passed through a cleaning autoencoder (top right), above a class activation map (bottom left), which highlights the regions of this spectrum that were important for the classification as a Goodenough structure. The neural network classifier now correctly infers that the input signal corresponds to the Goodenough model, with the class scores $[0, 1]$ corresponding to $[Y^{\mathrm{dimer}}, Y^{\mathrm{Goodenough}}]$. Looking at the activation maps, in particular the averages per slice of the input (bottom right), the classification is most heavily influenced by the signal in the slice corresponding to 35.6 to 40.4 meV in the spectrum collected at $E_i=70$~meV. This is the region at the top of the lower spin wave band and is precisely the same region as was used in the original paper to make the distinction. The activation map also has higher input from this same region at the top of the lower band in the spectra collected at $E_i=100$~and 140~meV, however this is not as strong as the contribution from the $E_i=70$~meV spectrum.  

These results demonstrate the importance of balancing complexity versus interpretability, when one is interested in understanding why a network provides the results that it does. In the case of the highly accurate network, which could take into account experimental noise we are able to obtain the correct result, however the layers of abstraction within the network make it difficult to explain these results in an interpretable way. When we construct a simpler network we have to remove some of the experimental noise in order to make it work properly, however the result can be interpreted in a way which corresponds to human understanding. This is an example of the well-known trade-off between interpretability and completeness of explanations \cite{gilpin2018explaining}.

We have implemented the CAM networks in an interactive \jupyter notebook~\cite{datarepo}.
In the notebook both the networks described above can be loaded and tested on the data before and after cleaning with the autoencoder. The notebook also features a tool for zooming in on slices of energy transfer space to look at the difference between dimer and Goodenough model spectra.

\section{Future challenges}

While we have addressed the classification problem in this work, another particular area of interest which we have not covered is the use of NNs for regression to obtain a set of exchange parameters for a given model and given measured data. With conventional fitting it might take a researcher some months (work in practice sometimes spread over years) to obtain a set of exchange parameters for a given model and measured dataset depending on the complexity of the data. This is due to the need first for a careful examination of the data for diagnostic regions and then a exhaustive search of the parameter space. Despite the required training time (perhaps a few weeks on a cluster), NNs offer a way to obtain these exchange parameter values on much shorter timescales.


The challenge in applying these functionalities (classification and regression) more generally are twofold. First, for users to take full advantage of the new methods, the workflows must be relatively easy to use and this will involve chaining together several disparate software packages in several languages. At the ISIS Neutron and Muon Source this will be addressed by the PACE project which aims to bring together data analysis tools for INS~\cite{PACE}. 

Second, as we noted in the introduction, one key feature of single-crystal INS data from modern spectrometers is their large size. The Pr(Ca,Sr)$_2$Mn$_2$O$_7$ dataset explored in this work is two orders of magnitude smaller than those typical of full 4D wave vector - energy transfer space maps of excitations, and yet we had to reduce it further to a 400$\times$240 pixel image for input to the neural network (the full data consist of 32 million $(\mathbf{Q}, \hbar\omega)$ bins although many of these will have zero counts). Thus INS data pose almost the opposite problem to that often tackled by machine learning, in that other applications can rely on a large number of datasets with each dataset being relatively small in size; for INS there are typically few \emph{experimental} datasets, necessitating the use of synthetic data for training, but each dataset is almost unmanageable in size. 

New techniques will have to be developed to address this challenge, and it is likely that machine learning methods will play a role. For example, although the total number of data bins is large, in many cases a significant fraction of these contain scattering that is not of relevance, for example, a slowly varying function of wave vector and energy transfer that forms a ``background'' underlying what is known \emph{a priori} to be sharply defined set of excitations of interest, such as the spin waves in the present work. If a reliable way can be found to identify, that is, segment this ''background'', then we can treat only the signal from the non-''background'' data. This signal might be amenable to Fourier analysis or to treatment using sum rules, or may be transformed into a compressed representation which might be more appropriate for input to a neural network.

\section{Conclusion}

We have applied deep neural networks to analyse INS data. This is the first time (to our knowledge) that NNs have been applied directly for INS data analysis. We demonstrated that a CNN architecture could be trained to classify spin Hamiltonians based on simulated data. We then demonstrated how uncertainty quantification can be included in the classification using the DUQ method. DUQ highlights that an accurate calculation of the experimental resolution is required to train a network using synthetic data that then works well on real data. DUQ was also able to detect when data from a different material to the training data but measured with the same spectrometer settings was passed to the network. 

We built class activation maps to determine which parts of the input spectrum are important for the network classification and found that while more complex networks can classify correctly with noisy data, the results are not physically interpretable. Simpler (fewer layer) networks, however, yield more interpretable class activation maps at the cost of requiring cleaner data which had been passed through an autoencoder. In this case, the maps showed that the most diagnostic feature of the measured spectrum is the region at the top of the lower excitation band, which accords with previous work~\cite{johnstone2012}.

We believe that this work demonstrates how trustable, interpretable ML can be a powerful tool in the analysis of INS data. We have highlighted future directions that can be explored to further harness the potential of ML in this field. We hope that this work is an early step in using ML methods not just for processing, but for better understanding INS experiments.

\section*{Acknowledgements}

Experiments at ISIS were supported by a beamtime allocation RB1310483 from the Science and Technology Facilities Council. TGP thanks co-authors A.T.Boothroyd and D.Prabhakaran of ref.~\cite{johnstone2012} for permission to use the datasets from those experiments. This work was partially supported by Wave 1 of The UKRI Strategic Priorities Fund under the EPSRC Grant EP/T001569/1, particularly the “AI for Science” theme within that grant and The Alan Turing Institute. The simulated datasets were generated using computing resources provided by STFC Scientific Computing Department's SCARF cluster.

\section*{Data Access Statement}

All of the training data, trained neural networks and code for generating the training data for this study are openly available at \url{https://zenodo.org/record/4270057#.X65xr1BpHIU}.

A git repository containing the code used to build and train the neural networks, as well as notebooks to recreate the DUQ and CAM experiments is available at \url{https://github.com/keeeto/interpretable-ml-neutron-spectroscopy}

The software environment required to run the codes in the git repository is available in the form of docker images from \url{https://hub.docker.com/u/mducle} with instructions in the \texttt{github} repository. Additionally there are Conda environment files provided in the repository.

\section*{Author Contributions}

KTB, MDL and TGP jointly conceived, planned and steered the project. KTB built, trained and applied the neural networks; MDL produced the simulated training datasets; TGP provided the experimental data. KTB, MDL and TGP wrote the manuscript together in an iterative fashion to ensure the contents present neural networks and related methods in a manner that is accessible and useful to the condensed matter physics community. JT was involved in conception and establishing of the project and facilitated the work in the paper.

\vspace{1cm}

\bibliographystyle{iopart-num}
\bibliography{mlrefs} 

\end{document}


\preprint{APS/123-QED}

\title{Interpretable neural networks for analysis and understanding of neutron spectra}

\author{Keith T. Butler}
\affiliation{SciML, Scientific Computing Department, Rutherford Appleton Laboratory, Harwell, OX110QX}
\email{keith.butler@stfc.ac.uk}
\author{Manh Duc Le}%
\affiliation{ISIS Neutron and Muon Source, Rutherford Appleton Laboratory, Harwell, OX110QX}
\author{Toby G. Perring}%
\affiliation{ISIS Neutron and Muon Source, Rutherford Appleton Laboratory, Harwell, OX110QX}

\date{\today}

\maketitle

\section{Model Training}
\begin{figure}
\centering
\begin{subfigure}
    \centering
    \includegraphics[width=0.85\columnwidth]{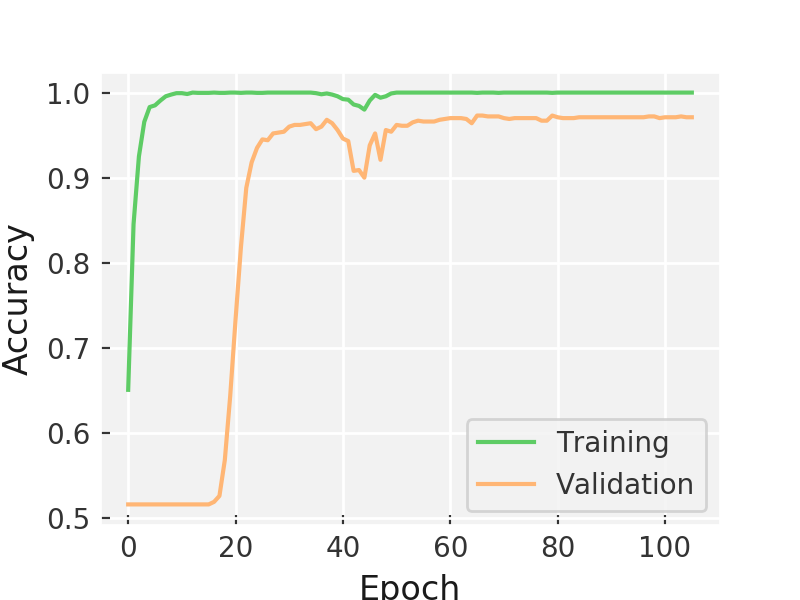}
\end{subfigure}
\begin{subfigure}
    \centering
    \includegraphics[width=0.85\columnwidth]{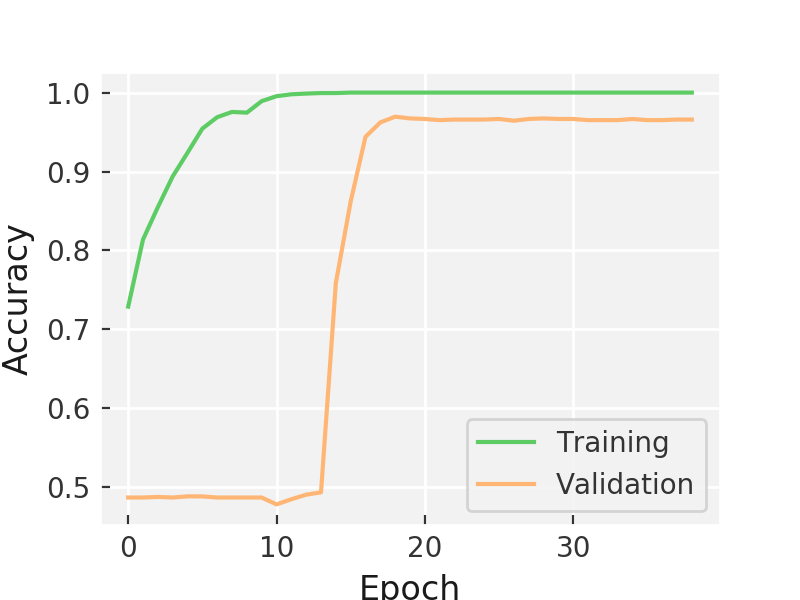}
\end{subfigure}

\caption{\label{fig:trainig} Upper: the training curve for the classification model trained on the data with pre-computed resolution functions. Lower: the training curve for the discrimination model trained on Monte Carlo calculated resolution functions.}
\end{figure}